\documentclass[12pt]{iopart}
 
\usepackage{iopams}
\usepackage{graphicx}

\newcommand{\Ne}[1]{\ensuremath{\mathcal{N}={#1}}} 
\newcommand{\ls}{\ensuremath{{l_{\rm s}}}} 
\newcommand{\gs}{\ensuremath{{g_{\rm s}}}} 
\newcommand{\hyp}[2]{\ensuremath{(\mathbf{#1},\mathbf{#2})}} 
\newcommand{\cZZ}[1]{{\ensuremath{\mathbb{C}^3/\mathbb{Z}_{#1}\times\mathbb{Z}_{#1}}}} 
\newcommand{\czz}{\cZZ{2}} 
\newcommand{\rc}{r_{\rm c}} 
\newcommand{\weff}{\ensuremath{{W_{\rm eff}}}}

\begin{document}

\begin{flushright}
{\small
ITP-UU-03/68\\
SPIN-03/44\\
hep-th/0312137\\ 
}
\end{flushright}

\title[$\mathcal{N}=1$ gauge superpotentials from supergravity]
{$\mathcal{N}=1$ gauge superpotentials from supergravity~\protect\footnote[1]{{\it Contribution to the proceedings of the workshop of the RTN Network ``The quantum structure of space-time and the geometric nature of fundamental interactions'', Copenhagen, September 2003.}}}

\author{Emiliano Imeroni}

\address{Institute for Theoretical Physics \& Spinoza Institute,\\ Utrecht University, Postbus 80.195, 3508 TD Utrecht, The Netherlands}

\ead{E.Imeroni@phys.uu.nl}

\begin{abstract}
We review the supergravity derivation of some non-perturbatively generated effective superpotentials for \Ne{1} gauge theories. Specifically, we derive the Veneziano--Yankielowicz superpotential for pure \Ne{1} Super Yang--Mills theory from the warped deformed conifold solution, and the Affleck--Dine--Seiberg superpotential for \Ne{1} SQCD from a solution describing fractional D3-branes on a \czz{} orbifold.
\end{abstract}

\section{Introduction and outlook}

The gauge/string theory correspondence, originally formulated for superconformal \Ne{4} Super Yang--Mills (SYM) theory, has in recent years been proven useful to yield relevant information on less supersymmetric and non conformal gauge theories too (see for instance~\cite{Herzog:2001xk} and references therein).

In order to build geometries dual to such more ``realistic'' gauge theories, one usually has to follow a two-step procedure:
\begin{enumerate}
	\item \emph{reduce supersymmetry} by choosing an appropriate closed string background, such as an orbifold or a Calabi--Yau manifold, known to break a certain amount of supersymmetry;
	\item \emph{break conformal invariance} by engineering suitable D-brane configurations (thus acting on the open string channel) with non-trivial world-volume topology.
\end{enumerate}

By following the above strategy, several string backgrounds were found, realizing for instance four-dimensional \Ne{1} and \Ne{2} SYM, such as configurations of fractional D3-branes on orbifolds and on the conifold, and of D5-branes wrapped on two-cycles inside Calabi--Yau manifolds.

In all these cases, low energy supergravity solutions describing the D-brane configurations have been used to succesfully extract some relevant pieces of information on the dual gauge theories, such as for instance:
\begin{itemize}
	\item \emph{at the perturbative level}, the running coupling constant, the chiral anomaly and the moduli space of the gauge theory;
	\item \emph{at the non-perturbative level}, the action of an instanton, realizations of chiral symmetry breaking and gaugino condensation, effective superpotentials and the tension of confining strings.
\end{itemize}

Here we concentrate on showing how it is possible to extract some non-perturbatively generated \emph{effective superpotentials for \Ne{1} gauge theories} starting from supergravity solutions and some geometric considerations.

A fundamental fact we will use is that \Ne{1} gauge theories can be realized in the framework of ``geometric transitions''~\cite{Gopakumar:1998ki,Vafa:2000wi,Cachazo:2001jy}, where one engineers them via configurations of D5-branes wrapped on two-cycles of resolved Calabi--Yau manifolds. The resulting geometry then flows to the one of a deformed manifold, where branes are replaced by fluxes.

In this context, the effective superpotential of the gauge theory is given in terms of the fluxes of the geometry by the following expression~\cite{Taylor:1999ii,Vafa:2000wi}:
\begin{equation}\label{Vafa}
	\weff \propto \sum_{i}
		\left[\ \int_{A_i} G_3 \int_{B_i} \Omega
		- \int_{A_i} \Omega \int_{B_i} G_3\ \right]\,.
\end{equation}
where $G_3 = \rmd C_2 + (C_0+\rmi\ \rme^{-\Phi})\ \rmd B_2$ is the complex three-form field strength of type IIB supergravity, $\Omega$ is the holomorphic $(3,0)$-form of the Calabi--Yau manifold, and $A_i$ and $B_i$ (which are respectively compact and non-compact) form a standard basis of orthogonal three-cycles on the manifold.

\section{VY superpotential from the conifold}

Let us start from the prototype example of the geometric transition framework, the \emph{conifold} (a cone over the Einstein manifold $T^{1,1}$), which is a singular non-compact Calabi--Yau threefold described by the equation $F(x,y,z,t)=0$ in $\mathbb{C}^4$, where:
\begin{equation}\label{singcon}
	F(x,y,z,t) = x^2 + y^2 + z^2 + t^2\,.
\end{equation}

\begin{figure}
\begin{center}
\includegraphics[scale=.6]{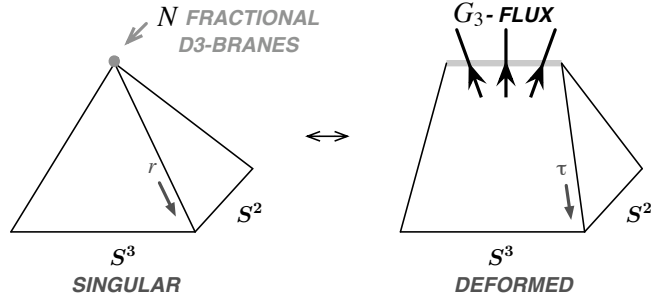}
\end{center}
\caption{Geometric transition of the conifold. Fractional D3-branes are replaced by $G_3$-flux through the three-cycles of the deformed manifold.\label{f:conifold}}
\end{figure}

In order to engineer pure \Ne{1} $SU(N)$ SYM, we put $N$ fractional D3-branes at the tip of the cone (these fractional branes have the interpretation of D5-branes wrapped on the two-cycle of the resolved manifold in the limit where its size shrinks to zero). The geometric transition brings us to the deformed conifold, which is a smooth manifold defined by replacing~\eref{singcon} with:
\begin{equation}\label{defcon}
	F(x,y,z,t) = x^2 + y^2 + z^2 + t^2 - \varepsilon^2\,,
\end{equation}
where $\varepsilon$ is a real positive deformation parameter. The fractional branes are replaced by the flux of $G_3$ through the newly blown-up compact three-cycle $A$ and through the non-compact three-cycle $B$, as shown in \fref{f:conifold}.

The corresponding supergravity solution is the \emph{warped deformed conifold}~\cite{Klebanov:2000hb}. Since we are going to use the solution for computing the fluxes needed in~\eref{Vafa}, we only need the expressions of $F_3=\rmd C_2$ and $B_2$ from~\cite{Klebanov:2000hb}:
\numparts
\begin{eqnarray}
	B_2 = \frac{\gs N\ls^2}{2}
		\left[k_-(\tau)\ g^1 \wedge g^2 + k_+(\tau)\ g^3 \wedge g^4 \right]\,,\label{B2KS}\\
	F_3 = \frac{\gs N\ls^2}{2} \left\{g^5 \wedge g^3 \wedge g^4
		+ d\left[ F(\tau) \left(g^1 \wedge g^3 + g^2 \wedge g^4\right)\right] \right\}\,,\label{F3KS}
\end{eqnarray}
\endnumparts
where $g^i$ are appropriate one-forms defined on the deformed conifold and where:
\begin{equation}
	k_{\pm}(\tau) = \frac{\tau\coth\tau-1}{2\sinh\tau}(\cosh\tau\pm 1)\,,\qquad
		F(\tau) = \frac{\sinh\tau-\tau}{2\sinh\tau}
\end{equation}
are functions of a dimensionless radial variable $\tau$. Let us consider the large $\tau$ limit, where we define a new radial coordinate $r$ such that $\tau \sim 3 \ln (r / r_0)$, $r_0$ being a regulator at short distances. In terms of $r$, the metric explicitly becomes the one of a cone over $T^{1,1}$, and we have~\cite{Klebanov:2000nc}:
\begin{equation}\label{BFKT}
	\fl B_2 \sim \frac{3\gs N\ls^2}{4} \ln\frac{r}{r_0}
		\left( g^1 \wedge g^2 + g^3 \wedge g^4 \right)\,,\qquad
	F_3 \sim \frac{\gs N\ls^2}{4} g^5 \wedge
		\left( g^1 \wedge g^2 + g^3 \wedge g^4 \right)\,.
\end{equation}
From~\eref{BFKT}, we can compute the fluxes of $G_3$ through the cycles $A$ and $B$ defined by:
\begin{equation}
	\fl A:\qquad g^1=-g^3\,,\quad g^2=-g^4\,,\quad r\ {\rm constant}\,;\qquad
	B:\qquad g^1=g^2=g^5=0\,.
\end{equation}
Using the fact that $\int_A g^3\wedge g^4\wedge g^5 = 8\pi^2$ and $\int_B \rmd r \wedge g^3\wedge g^4 = 8\pi \int \rmd r$, we get:
\begin{equation}\label{Gflux}
	\frac{1}{(2\pi\ls)^2\gs}\int_A G_3 = N\,,\qquad
	\frac{1}{(2\pi\ls)^2\gs}\int_B G_3 = - \frac{3N}{2\pi \rmi} \ln \frac{\rc}{r_0}\,,
\end{equation}
where we had to introduce a cut-off $\rc$ in order to perform the integration over $r$.%
\footnote[2]{Using the asymptotic expressions~\eref{BFKT} does not affect the correctness of the result~\eref{Gflux}, since it is precisely the identification of the fluxes between the full and asymptotic solutions which is used to derive the relation between the coordinates $\tau$ and $r$.}

Now we need to compute the periods of the holomorphic $(3,0)$-form:
\begin{equation}
	\Omega = \frac{\rmd x\wedge \rmd y\wedge \rmd z}{2\sqrt{\varepsilon^2-x^2-y^2-z^2}}\,.
\end{equation}
The computation (see for example~\cite{Cachazo:2001jy}) reduces to the one of the one-dimensional integral $\int \rmd x\ \sqrt{\varepsilon^2-x^2}$, where the extrema of integration are given by $x\in[-\varepsilon,\varepsilon]$ for the $A$-cycle, and by $x\in[\varepsilon,\rc^{3/2}]$ for the $B$-cycle. Notice that we are using the same cut-off used in the calculation of the fluxes of $G_3$, and that the power $\rc^{3/2}$ is due to the fact that the coordinates in~\eref{defcon}, as well as $\varepsilon$, have dimensions of the $3/2$ power of a length, as can be seen from the full metric in~\cite{Klebanov:2000hb}. The periods of $\Omega$ then read:
\begin{equation}\label{periods}
	\fl\int_A \Omega = 2\pi\rmi\ \frac{\varepsilon^2}{4}\,,\qquad
	\int_B \Omega = \frac{1}{2i} \left(
		\rc^{3/2}\sqrt{\varepsilon^2-\rc^{3}}
		+\varepsilon^2\arcsin\frac{\rc^{3/2}}{\varepsilon}-\frac{\pi\varepsilon^2}{2}\right)\,,
\end{equation}
and we can expand the $B$-period for large $\rc$, getting:
\begin{equation}\label{Bperiod}
	\int_B \Omega \sim \frac{\rc^3}{2}-\frac{\varepsilon^2}{4}
		+\frac{\varepsilon^2}{4}\ln\frac{\varepsilon^2}{4\rc^3}\,.
\end{equation}

We can now substitute the results~\eref{Gflux}, \eref{periods}, \eref{Bperiod} inside the formula~\eref{Vafa}, obtaining the following expression for the effective gauge superpotential in terms of supergravity quantities:
\begin{equation}\label{sugraW}
	\weff \propto -\frac{N\varepsilon^2}{4}\left(1-\ln\frac{\varepsilon^2}{4r_0^3}\right) + \frac{N\rc^3}{2}\,.
\end{equation}

We still have to express~\eref{sugraW} in terms of gauge theory quantities. This can be done by implementing the ``stretched string'' energy/radius relations:
\begin{equation}
	\rc = 2\pi\ls^2\ \mu\,,\qquad
	r_0 = 2\pi\ls^2\ \Lambda\,,\qquad
	\frac{\varepsilon^2}{4} = (2\pi\ls^2)^3\ S\,,
\end{equation}
where $\mu$ is the subtraction scale of the gauge theory, $\Lambda$ is the dynamical scale, and $\varepsilon^2$ is naturally identified, due to its engineering dimensions, with the gaugino condensate $S$ of the dual gauge theory.

Reinstating the correct units in~\eref{sugraW}, neglecting the unimportant term independent of $S$ and using the above energy/radius relations, we finally get the following answer (first derived in this setup in~\cite{Cachazo:2001jy,Giddings:2001yu}) for the effective superpotential:
\begin{equation}\label{VY}
	\weff = NS \left(1-\ln\frac{S}{\Lambda^3}\right)\,.
\end{equation}
This is the well-known Veneziano--Yankielowicz superpotential for pure \Ne{1} SYM~\cite{Veneziano:1982ah}. We have therefore seen that supergravity, together with some geometric considerations, is able to give us the correct answer for the non-perturbatively generated effective superpotential of the dual gauge theory.%
\footnote[3]{Notice that, in the computation of the superpotential, we could use the full expression in~\eref{periods} of the $B$-period of $\Omega$ instead of its large $\rc$ expression~\eref{Bperiod}. Doing so would result in corrections to the superpotential which look like fractional instanton contributions, analogous to the ones found for the running coupling constant of pure \Ne{1} SYM in~\cite{DiVecchia:2002ks}.}

\section{SQCD moduli space and ADS superpotential from fractional branes}

We now want to engineer SQCD, namely an \Ne{1} gauge theory with fundamental matter. To achieve this goal, let us consider a system of fractional D3-branes transverse to a \czz{} orbifold~\cite{Douglas:1997de}. This orbifold, which breaks bulk supersymmetry down to eight supercharges, is defined by the action on the complex coordinates $z_k = x_{2k+2}+\rmi\ x_{2k+3}$ of the generators of the two $\mathbb{Z}_2$ factors:
\begin{equation}
	\fl g_1:\quad z_2\to-z_2\,,\quad z_3\to-z_3\,,\qquad
	g_2:\quad z_1\to-z_1\,,\quad z_3\to-z_3\,.
\end{equation}

Fractional branes are the most elementary brane objects on orbifolds~\cite{Douglas:1996sw}. They are defined by the fact that the Chan--Paton factors of the string attached to them transform in irreducible representations of the orbifold group. Since \czz{} has four irreducible one-dimensional representations, we will have four different types of fractional branes, that we denote with A, B, C, D. They are charged under the open string twisted sectors, and they are stuck at the orbifold fixed point $z_i=0$.

\begin{figure}
\begin{center}
\includegraphics[scale=.7]{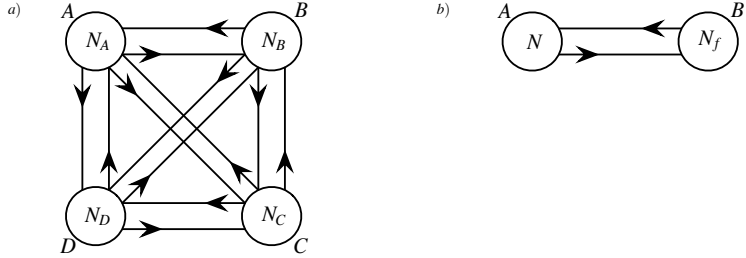}
\end{center}
\caption{Quiver diagrams for fractional D-branes on \czz. $a)$ The full diagram. Each node $i$ represents a gauge group factor $U(N_i)$ with the corresponding vector multiplet, while each oriented arrow from node $i$ to node $j$ represents a chiral multiplet in the \hyp{N_i}{\bar{N}_j} representation. $b)$ The diagram of the system of D-branes which realizes $U(N)$ SQCD with $N_f$ flavours on the branes of type A.\label{f:quivers}}
\end{figure}

The low-energy theory living on a generic system of fractional D3-branes of all types is a four-dimensional \Ne{1} gauge theory with gauge group $U(N_A)\times U(N_B)\times U(N_C)\times U(N_D)$ and twelve bifundamental chiral multiplets, whose quiver diagram is depicted in~\fref{f:quivers}a. If we take $N$ branes of type A and $N_f$ of type B (and no other branes, thus realizing the quiver diagram of~\fref{f:quivers}b), and concentrate only on the low-energy degrees of freedom living on the branes of type A, we see that we obtain \Ne{1} $U(N)$ SQCD with $N_f$ fundamental flavours of ``quarks'' $Q^i$ and ``antiquarks'' $\tilde{Q}_{\tilde{\jmath}}$~\cite{Berenstein:2003fx}. In what follows, we will concentrate on the case $N_f < N$, namely the Affleck--Dine--Seiberg theory~\cite{Affleck:1983mk}.

Let us first notice that the brane configuration under study encodes very naturally information about the classical moduli space of the gauge theory. The main observation is that the fractional branes of type A and B have the same charge under the sectors twisted by $g_2$ and $g_3=g_1g_2$, but opposite charge under the sector twisted by $g_1$. This means that a superposition A$+$B is only charged under this latter twisted sector, and is therefore no longer constrained to stay at the orbifold fixed point. Rather, it can freely move in the $z_1$ plane (provided of course that we introduce brane images on the covering space in order to make the configuration invariant under the orbifold action).

\begin{figure}
\begin{center}
\includegraphics[scale=.6]{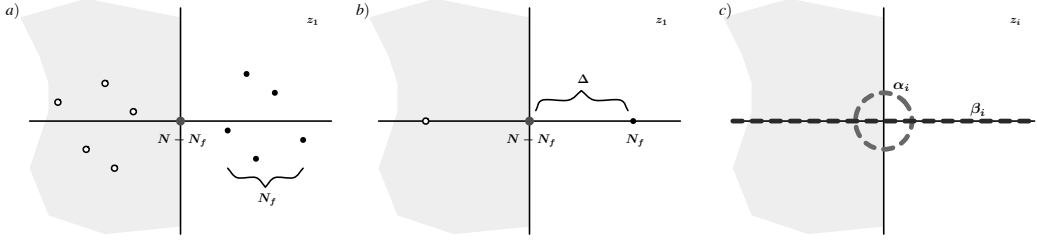}
\end{center}
\caption{Moduli space of the ADS theory via fractional branes. $a)$ A$+$B superpositions at arbitrary points of the $z_1$ plane. $b)$ The configuration which makes the meson matrix proportional to the identity. $c)$ The compact 1-cycle $\alpha_i$ and the non-compact one $\beta_i$ in the $z_i$ plane.\label{f:ads}}
\end{figure}

This fact gives a natural geometrical meaning to the moduli space of \Ne{1} SQCD for $N_f<N$, as shown in~\fref{f:ads}a. In fact, we can form $N_f$ A$+$B superpositions, and place them at arbitrary points in the $z_1$ plane, while the remaining $N-N_f$ fractional branes of type A are still stuck at the origin. This means that the low-energy effective description of the theory is in terms of a $U(N-N_f)$ gauge theory together with arbitrary expectation values of the meson matrix $M^i_{\phantom{i}\tilde{\jmath}}=Q^i\tilde{Q}_{\tilde{\jmath}}$, which in our case are mapped to the arbitrary positions of the A$+$B superpositions. Without loss of generality, we can make the meson matrix proportional to the identity, $M^i_{\phantom{i}\tilde{\jmath}} = m^2 \delta^i_{\phantom{i}\tilde{\jmath}}$, by putting all superpositions at a point $\Delta$ of the real axis of the $z_1$ plane, where $\Delta=2\pi\ls^2\ m$, as in~\fref{f:ads}b.

We would now like to pass to the computation of the non-perturbative effective superpotential. The (singular) supergravity solution describing fractional D3-branes on \czz{} was found in~\cite{Bertolini:2001gg}. Again, we only need the explicit expression of $G_3$, which in the case of the configuration depicted in~\fref{f:ads}b is:
\begin{equation}\label{G3orb}
	G_3 = d\gamma_i \wedge \omega^i\,,
\end{equation}
where $\omega^i$, $i=1,2,3$, are the anti-self dual $(1,1)$-forms dual to the shrinking 2-cycles $\mathcal{C}_i$ of the orbifold geometry and:
\begin{equation}\label{gamma}
        \fl\gamma_i = 4\pi\rmi\gs\ls^2\ \left[ (N - N_f) \ln \frac{z_i}{\epsilon_0}
                + \delta_{i,1}\ N_f\ \ln \frac{z_1-\Delta}{\epsilon_0}
                + \delta_{i,1}\ N_f\ \ln \frac{z_1+\Delta}{\epsilon_0} \right]\,,
\end{equation}

In order to use~\eref{Vafa}, we need to identify the appropriate three-cycles $A_i$ and $B_i$ in our orbifold geometry. We can introduce them by simply taking the direct product of the two-cycles $\mathcal{C}_i$ with suitable one-cycles on the $z_i$ planes. Specifically, we define:
\begin{equation}\label{ABcycles}
        A_i = \alpha_i \times \mathcal{C}_i\,,\qquad
        B_i = \beta_i \times \mathcal{C}_i\qquad (i=1,2,3)\,,
\end{equation}
where the compact cycles $\alpha_i$ and the non-compact cycles $\beta_i$ in the $z_i$ plane are orthogonal to each other and are shown in~\fref{f:ads}c. We now have all the necessary ingredients to compute the fluxes of $G_3$:
\begin{equation}\label{fluxorb}
	\fl\frac{1}{4\pi\rmi\gs\ls^2} \int_{A_i} G_3 = 2\pi \rmi\ ( N - N_f )\,,\quad
	\frac{1}{4\pi\rmi\gs\ls^2} \int_{B_i} G_3 \sim ( N - N_f )\ \ln \frac{r_c}{r_0}
		+ \delta_{i,1}\ 2 N_f\ \ln \frac{\Delta}{\rho_0}\,,
\end{equation}
where, similarly to the case considered in the previous section, we have introduced long and short distance cut-offs $\rc$ and $r_0$ in order to perform the integration over the non-compact cycles $B_i$. Notice also that we have expanded the result assuming $r_0 \ll r_c \ll \Delta$.

The next step is to compute the periods of the holomorphic $(3,0)$-form $\Omega$. The \czz{} orbifold is defined by the equation $F(x,y,z,t) = 0$ in $\mathbb{C}^4$, where:
\begin{equation}\label{singorb}
	F(x,y,z,t) = xyz + t^2\,.
\end{equation}
In order to de-singularize this space, we again introduce a positive deformation parameter $\xi$~\cite{Berenstein:2003fx}, and replace~\eref{singorb} by:
\begin{equation}\label{deforb}
	F(x,y,z,t) = xyz + t^2 - \xi^2\,.
\end{equation}
In this deformed orbifold, the periods of
\begin{equation}
	\Omega = \frac{\rmd x \wedge \rmd y \wedge \rmd z}{2\sqrt{\xi^2-xyz}}
\end{equation}
can be computed in a way analogous to the one followed in the previous section, and one gets:
\begin{equation}\label{perorb}
	\frac{1}{16\pi^2} \int_{A_i} \Omega = \xi\,,\qquad
	\frac{1}{16\pi^2} \int_{B_i} \Omega
		= -\frac{1}{2\pi \rmi}\ \frac{\xi}{3}\ \ln \frac{\xi}{\rho_c^3}\,.
\end{equation}

By substituting~\eref{fluxorb} and~\eref{perorb} in~\eref{Vafa}, and implementing again the ``stretched string'' energy/radius relations, which now read:
\begin{equation}
	\fl\rc = 2\pi\ls^2\ \mu\,,\qquad
	r_0 = 2\pi\ls^2\ \Lambda\,,\qquad
	\Delta = 2\pi\ls^2\ m\,,\qquad
	\xi = (2\pi\ls^2)^3\ S\,,
\end{equation}
we get the following result for the effective superpotential of \Ne{1} $U(N)$ SQCD with $N_f<N$ fundamental flavours:
\begin{equation}\label{VYT}
	\weff = ( N - N_f )\ \left[\ S - S\ 
		\ln \frac{S}{\Lambda^3}\ \right]
		- S\ \ln \frac{m^{2 N_f}}{\Lambda^{2 N_f}}\,.
\end{equation}
This is precisely the expected expression of the Taylor--Veneziano--Yankielowicz superpotential~\cite{Taylor:1983bp}. Minimizing with respect to $S$, we are finally left with the Affleck--Dine--Seiberg superpotential~\cite{Affleck:1983mk}:
\begin{equation}\label{ADS}
	\weff  = ( N - N_f )\ \left[\
		\frac{\Lambda^{3N-N_f}}{m^{2N_f}}\ \right]^{\frac{1}{N-N_f}}
                = ( N - N_f )\ \left[\ \frac{\Lambda^{3N-N_f}}{\det M}\ \right]^{\frac{1}{N-N_f}}
\end{equation}
(recall that we are working with a diagonal meson matrix so that $\det M = m^{2 N_f}$).

Some final comments are in order. First, notice that our brane construction is valid also for $N_f=0$ and $N_f=N$. In the former case, we recover again the Veneziano--Yankielowicz superpotential~\eref{VY}, while in the latter case our result~\eref{VYT} correctly reproduces the quantum constraint $\det M = \Lambda^{2N}$.

On the other hand, for $N_f>N$ our construction breaks down, which is a signal that one should incorporate Seiberg duality~\cite{Seiberg:1994pq} into the picture. An investigation of this issue in a related setup was performed in~\cite{Berenstein:2002fi} with formal methods, but a full understanding in terms of a supergravity dual is still lacking.

\ack
I wish to thank M.~Bertolini, M.~Bill\`o, P.~Di Vecchia, P.~Merlatti, G.~Vallone and in particular A.~Lerda for collaboration and discussions. Work partially supported by the European Commission RTN programme HPRN-CT-2000-00131.

\section*{References}

\providecommand{\href}[2]{#2} 
\end{document}